\documentclass[compress,preprint,12pt]{elsarticle}
\pdfoutput=1
\usepackage[utf8x]{inputenc}
\usepackage{microtype}
\usepackage{amsmath}
\usepackage{amssymb}
\usepackage{todonotes}
\usepackage{graphicx}
\usepackage{xcolor}
\usepackage{listings}
\usepackage{xspace}
\usepackage{hyperref}

\newenvironment{absolutelynopagebreak}
  {\par\nobreak\vfil\penalty0\vfilneg
   \vtop\bgroup}
  {\par\xdef\tpd{\the\prevdepth}\egroup
   \prevdepth=\tpd}

\DeclareRobustCommand{\ensuremathrm}[1]{\ensuremath{\mathrm{#1}}\xspace}

\DeclareRobustCommand{\HEJ}{\ensuremathrm{HEJ}}
\DeclareRobustCommand{\HIGHEJ}{\emph{High Energy Jets}}

\newcounter{bla}

\journal{Computer Physics Communications}

\begin{document}

\begin{absolutelynopagebreak}
\begin{frontmatter}
\noindent DESY 19-028\\
IPPP/19/13\\
MCnet-19-05\\

\title{HEJ 2: High Energy Resummation for Hadron Colliders}

\author[a]{Jeppe~R.~Andersen}
\author[a]{Tuomas~Hapola}
\author[a]{Marian~Heil}
\author[b]{Andreas~Maier\corref{author}}
\author[c]{Jennifer~Smillie}

\cortext[author] {Corresponding author.\\\textit{E-mail address:} andreas.martin.maier@desy.de}
\address[a]{Institute for Particle Physics Phenomenology,\\University of
  Durham, South Road, Durham DH1 3LE, UK}
\address[b]{Deutsches Elektronen-Synchrotron, DESY, Platanenallee 6,
  15738 Zeuthen, Germany}
\address[c]{Higgs Centre for Theoretical Physics, University of Edinburgh,\\
  Peter Guthrie Tait Road, Edinburgh EH9 3FD, UK}

\begin{abstract}
  We present HEJ 2, a new implementation of the \HIGHEJ{} formalism for
  high-energy resummation in hadron-collider processes as a flexible Monte
  Carlo event generator. In combination with a conventional fixed-order
  event generator, HEJ 2 can be used to obtain greatly improved
  predictions for a number of phenomenologically important processes by
  adding all-order logarithmic corrections in $\hat{s}/p_\perp^2$. A prime
  example for such a process is the gluon-fusion production of a Higgs
  boson in association with widely separated jets, which constitutes the
  dominant background to Higgs boson production in weak-boson fusion.
\end{abstract}


\end{frontmatter}
\end{absolutelynopagebreak}
{\bf PROGRAM SUMMARY}

\begin{small}
\noindent
{\em Program Title:} HEJ 2                                     \\
{\em Licensing provisions:} GPLv2 or later                             \\
{\em Programming language:} C++                                \\
{\em Nature of problem:}\\ 
Conventional event generators for high-energy colliders are based on
fixed-order perturbation theory, which converges poorly in regions of
phase space with large invariant masses between jets. However, reliable
predictions in these regions are of particular importance whenever cuts
relevant to  weak-boson fusion or weak-boson scattering are applied. One highly relevant
example is the separation of gluon fusion and weak-boson fusion
contributions to the production of a Higgs boson in association with two
or more jets.\\
{\em Solution method:}\\ 
  The perturbative description can be amended by resumming the
  high-energy logarithms associated with large invariant masses to all
  orders in perturbation theory. HEJ 2 implements this resummation
  following the \HIGHEJ{} framework. Using HEJ 2 in conjunction with a
  conventional event generator allows one to obtain reliable predictions for
  processes involving multiple jets in high-energy regions of phase space.\\
{\em Additional comments including Restrictions and Unusual features:}\\
The current version of HEJ 2 implements multi-jet processes with at most
a single Higgs boson in the final state. Matching to fixed-order
perturbation theory is currently restricted to leading order.
   \\


\end{small}

\section{Introduction}
\label{sec:intro}

Measurements at the Large Hadron Collider (LHC) are testing our
know\-ledge of elementary particle physics to unprecedented
detail. Especially in the light of the High-Luminosity LHC upgrade it is
instrumental to develop a thorough understanding of scattering processes
over a wide range of phase space, including specific corners of
interest.

The high-energy region of phase space is characterised by large
invariant masses, or alternatively large rapidity separations, between pairs of
jets. This region is of particular relevance in the production of a
Higgs boson together with jets through weak-boson fusion, where
the partons are typically scattered in the forward region. Similar
kinematics are observed in weak-boson scattering. Experimental
identification of such processes requires a precise and reliable
theoretical prediction for the background, which is dominated by gluon
fusion.

In the high-energy limit, the convergence of conventional perturbation
theory is spoiled by large logarithms of invariant masses. While these
logarithms can be predicted and resummed to all orders in the framework
of BFKL\cite{Fadin:1975cb,Kuraev:1976ge,Kuraev:1977fs,Balitsky:1978ic},
it is found that a prediction based on \emph{only} the logarithmic
behaviour is generally not sufficient in the phase space probed by
current colliders. The \HIGHEJ{} framework pioneered
in~\cite{Andersen:2009nu,Andersen:2009he,Andersen:2011hs} remedies this
shortcoming by
performing a systematic power expansion of the scattering amplitude thereby
retaining the logarithmic accuracy of BFKL while allowing event-by-event matching to
fixed-order matrix elements.
As approximations are only applied to the scattering
matrix elements and not to the phase space, it is possible to obtain
fully exclusive predictions. A review of \HIGHEJ{} can be found
in~\cite{Andersen:2017kfc}. The method was implemented in the \HEJ{}
Monte Carlo event generator and validated by comparing to measurements
of multi-jet production and the production of $W$ or $Z$ bosons in
associations with jets in the high-energy
region~\cite{Andersen:2012gk,Andersen:2016vkp,Aad:2011jz,Chatrchyan:2012gwa,Chatrchyan:2012pb,Abazov:2013gpa,Aad:2014pua,Aad:2014qxa}.

In order to benefit from recent progress in the formulation of
\HIGHEJ{}, it has been necessary to rewrite large parts of the
implementation. Here, we present the new program HEJ 2. HEJ 2 is based
on a recently developed matching algorithm, where events in the
resummation phase space are generated from fixed-order events produced
with a conventional fixed-order event generator. A detailed description
of the algorithm and its implementation is given
in~\cite{Andersen:2018tnm}. In contrast to the previously used
formulation, it allows fixed-order matching to much higher
multiplicities and to loop-induced processes.  As matching is no longer tied to
hard-wired matrix elements, the scope of matching is much broader
than with the previous version of HEJ.

HEJ 2 has a much more modular structure, which makes it easier to add
and maintain code for new processes. It is straightforward to interface
HEJ 2 to arbitrary fixed-order event generators producing events in the
Les Houches Event File format. HEJ 2 natively supports
\texttt{Rivet}~\cite{Buckley:2010ar} analyses and the standard
\texttt{LHEF}~\cite{Boos:2001cv,Alwall:2006yp} and
\texttt{HepMC}~\cite{Dobbs:2001ck,HepMC} formats for event output. It is
also possible to use custom analyses. Not only can HEJ 2 be used as a
stand-alone program, but also as a library. This will for example
facilitate future combinations with parton showers,
cf.~\cite{Andersen:2011zd,Andersen:2017sht}.

In the following we explain how to install and run HEJ 2 in
sections~\ref{sec:install} and \ref{sec:HEJ2_run},
respectively. Section~\ref{sec:HEJFOG} describes the HEJ fixed-order
generator, which can generate leading-order high-multiplicity events in
the high-energy approximation. In sections~\ref{sec:custom_analysis} and
~\ref{sec:scales}, we show how a user-defined analysis and custom
renormalisation and factorisation scales can be defined. Finally, in
section~\ref{sec:lib_use} we give an example how HEJ 2 can be used as a
library. Continuously updated documentation for HEJ can be found at \url{https://hej.web.cern.ch}.


\section{Installation}
\label{sec:install}

\subsection{Download}
\label{download}

A tar archive of the HEJ 2 source code can be downloaded and
decompressed with the command
\begin{lstlisting}[language=sh]
curl https://hej.web.cern.ch/HEJ/downloads/HEJ_2.0.tar.gz | tar -xz
\end{lstlisting}
To obtain the latest stable HEJ version, \lstinline!HEJ_2.0.tar.gz!
should be replaced by \lstinline!HEJ.tar.gz!.

Alternatively, the HEJ source code can be obtained by installing the
git version control system~\cite{git} and running
\begin{lstlisting}[language=sh]
git clone https://phab.hepforge.org/source/hej.git
\end{lstlisting}
We also provide a Docker~\cite{docker} image containing a HEJ 2
installation on \url{https://hub.docker.com/r/hejdock/hej}. This image
can be pulled with
\begin{lstlisting}
docker pull hejdock/hej
\end{lstlisting}
When using the Docker image the remaining installation steps can be
skipped.

\subsection{Prerequisites}
\label{sec:prereq}

Before installing HEJ 2, the following programs and libraries are needed:
\begin{itemize}
\item CMake version 3.1~\cite{Martin:2007,cmake} or later.
\item  A compiler supporting the C++14 standard, for
  example gcc 5~\cite{gcc} or later.
\item FastJet~\cite{Cacciari:2011ma}.
\item CLHEP version 2.3.1 or later~\cite{CLHEP}
\item LHAPDF6~\cite{Buckley:2014ana,LHAPDF}.
\item The IOStreams and uBLAS boost libraries~\cite{boost}.
\item yaml-cpp~\cite{yaml-cpp}.
\end{itemize}
HEJ~2 was tested with CMake version 3.12.3, gcc 5.3.0, FastJet 3.3.1,
CLHEP 2.4.1.0, LHAPDF 6.1.6, boost 1.68, and yaml-cpp 0.6.2 as well as
with various other combinations. For including finite top mass corrections in
Higgs boson plus jets production, version 2 of the QCDLoop
library~\cite{Carrazza:2016gav,qcdloop} is required in addition to the
above libraries. HEJ 2 supports \texttt{Rivet}~\cite{Buckley:2010ar} and versions 2 and 3 of
\texttt{HepMC}~\cite{Dobbs:2001ck,HepMC} if any of them are installed,
but does not require them.

\subsection{Compilation}
\label{sec:compile}

HEJ 2 can be compiled and installed with the commands
\begin{lstlisting}[language=sh]
cmake HEJ2/source/directory -DCMAKE_INSTALL_PREFIX=target/directory
make install
\end{lstlisting}
where \lstinline!HEJ2/source/directory! is the directory containing the file
\lstinline!CMakeLists.txt!. If \lstinline!-DCMAKE_INSTALL_PREFIX=target/directory! is omitted
HEJ 2 will be installed to some default location.

In case some of the aforementioned prerequisites are not found by cmake
a hint can be given by adding an additional argument
\begin{lstlisting}[language=sh]
-Dlibname_ROOT_DIR=/directory/with/library
\end{lstlisting}
 where \lstinline!libname! should be
replaced by the name of the library in question.

\subsection{Testing}
\label{sec:test}

The installation can be tested after downloading the NNPDF 2.3 PDF set with
\begin{lstlisting}[language=sh]
lhapdf install NNPDF23_nlo_as_0119
\end{lstlisting}
The tests can be run with the command
\begin{lstlisting}
make test
\end{lstlisting}


\section{Running HEJ 2}
\label{sec:HEJ2_run}

\subsection{Quick start}
\label{sec:start}

In order to run, HEJ 2 needs a configuration file and a file containing
fixed-order events. A sample configuration is given by the
\lstinline!config.yml! file distributed together with HEJ 2. Events in
the Les Houches Event File format can be generated with standard Monte
Carlo generators like \texttt{MadGraph5\_aMC@NLO}~\cite{Alwall:2014hca}
or \texttt{Sherpa}~\cite{Gleisberg:2008ta}. It is also possible to use
Les Houches Event Files compressed with \texttt{gzip}~\cite{gzip} as
input. HEJ 2 assumes that the cross section is given by the sum of the
event weights. Depending on the fixed-order generator it may be
necessary to adjust the weights in the Les Houches Event File
accordingly.

The processes supported by HEJ 2 so far are
\begin{itemize}
\item pure multijet production,
\item production of a Higgs boson with jets,
\end{itemize}
where at least two jets are required in each case. Only leading-order
events are supported; work is ongoing to extend the matching to next-to-leading-order.

After generating an event file \lstinline!events.lhe!, adjust the parameters
under the \lstinline!fixed order jets! setting in \lstinline!config.yml! to the
settings in the fixed-order generation. Resummation can then be added by
running
\begin{lstlisting}
HEJ config.yml events.lhe
\end{lstlisting}
Using the default settings, this will produce an output event file
\lstinline!HEJ.lhe! with events including high-energy resummation.

When using the Docker image, HEJ can be run with
\begin{lstlisting}[language=bash]
docker run -v $PWD:$PWD -w $PWD hejdock/hej HEJ config.yml events.lhe
\end{lstlisting}

\subsection{Settings}
\label{sec:settings}

HEJ 2 configuration files follow the YAML~\cite{YAML} format. The
following configuration parameters are supported:

\begin{itemize}
\item \textbf{trials}: High-energy resummation is performed by
  generating a number of resummation phase space configurations
  corresponding to an input fixed-order event. This parameter specifies
  how many such configurations HEJ 2 should try to generate for each
  input event. Typical values vary between 10 and 100.

\item   \textbf{min extparton pt}: Specifies the minimum transverse momentum
  in GeV of the most forward and the most backward parton. This setting
  is needed to regulate an otherwise uncancelled divergence. Its value
  should be slightly below the minimum transverse momentum of jets
  specified by \textbf{resummation jets: min pt}. See also the \textbf{max ext soft
  pt fraction} setting.

\item   \textbf{max ext soft pt fraction}: Specifies the maximum fraction that
  soft radiation can contribute to the transverse momentum of each the
  most forward and the most backward jet. Values between around 0.05 and
  0.1 are recommended. See also the \textbf{min extparton pt} setting.

\item   \textbf{fixed order jets}: This tag collects a number of settings
  specifying the jet definition in the event input. The settings should
  correspond to the ones used in the fixed-order Monte Carlo that
  generated the input events.

  \begin{itemize}
  \item \textbf{min pt}: Minimum transverse momentum in GeV of
    fixed-order jets.

  \item \textbf{algorithm}: The algorithm used to define jets. Allowed
    settings are \lstinline!kt!, \lstinline!cambridge!, \lstinline!antikt!, \lstinline!cambridge for passive!.
    See the FastJet documentation for a description of these algorithms~\cite{Cacciari:2011ma}.

  \item \textbf{R}: The R parameter used in the jet algorithm, roughly
    corresponding to the jet radius in the plane spanned by the rapidity
    and the azimuthal angle.
  \end{itemize}

\item \textbf{resummation jets}: This tag collects a number of settings
  specifying the jet definition in the observed, i.e. resummed
  events. These settings are optional, by default the same values as for
  the \textbf{fixed order jets} are assumed.

  \begin{itemize}
  \item \textbf{min pt}: Minimum transverse momentum in GeV of
    resummation jets. This should be between
    25\% and 50\% larger than the minimum transverse
    momentum of fixed order jets set by \textbf{fixed order jets: min pt}.

  \item \textbf{algorithm}: The algorithm used to define jets. HEJ 2
    can cover the resummation phase space particularly efficiently when using
    \lstinline!antikt! jets~\cite{Cacciari:2008gp},
    so this value is strongly recommended. For a list of possible other
    values, see the \textbf{fixed order jets: algorithm} setting.

  \item \textbf{R}: The R parameter used in the jet algorithm.
  \end{itemize}
\item \textbf{FKL}: Specifies how to treat events respecting FKL
  rapidity ordering. These configurations are dominant in the high-energy
  limit (see~\cite{Andersen:2017kfc}). The possible values are
  \lstinline!reweight! to enable resummation, \lstinline!keep! to keep the
  events as they are up to a possible change of renormalisation and
  factorisation scale, and \lstinline!discard! to discard these events.

\item \textbf{unordered}: Specifies how to treat events with one
  emission that does not respect FKL ordering. In the high-energy limit,
  such configurations are logarithmically suppressed compared to FKL
  configurations. The possible values are the same as for the \textbf{FKL}
  setting, but \lstinline!reweight! is currently only supported for
  Higgs boson plus jets production.

\item   \textbf{non-HEJ}: Specifies how to treat events where no resummation
  is possible. The allowed values are \lstinline!keep! to keep the events as
  they are up to a possible change of renormalisation and factorisation
  scale and \lstinline!discard! to discard these events.

\item   \textbf{scales}: Specifies the renormalisation and factorisation
  scales for the output events. This can either be a single entry or a
  list \lstinline![scale1, scale2, ...]!. For the case of a list the first
  entry defines the central scale. Possible values are fixed numbers to
  set the scale in GeV or the following:

  \begin{itemize}
  \item \lstinline!H_T!: The sum of the scalar transverse momenta of
    all final-state particles.
  \item \lstinline!max jet pperp!: The maximum
    transverse momentum of all jets.
  \item \lstinline!jet invariant mass!:
    Sum of the invariant masses of all jets.
  \item \lstinline!m_j1j2!: Invariant mass between the two hardest jets.
  \end{itemize}
  Scales can be multiplied or divided by an overall factor,
  e.g. \lstinline!H_T/2!.

  It is also possible to import scales from an external library, see
  section~\ref{sec:scales}.

\item \textbf{scale factors}: A list of numeric factors by which each of
  the scales should be multiplied. The renormalisation scale $\mu_r$ and
  the factorisation cales $\mu_f$ are varied independently. For example, a
  list with entries \lstinline![0.5, 2]!  would give the four scale
  choices $(0.5\mu_r, 0.5\mu_f)$, $(0.5\mu_r, 2\mu_f)$, $(2\mu_r, 0.5\mu_f)$,
  $(2\mu_r, 2\mu_f)$ in this order. The ordering corresponds to the order
  of the final event weights.

\item   \textbf{max scale ratio}: Specifies the maximum factor by which
  renormalisation and factorisation scales may differ. For a value of
  $2$ and the example given for the scale factors the scale
  choices $(0.5\mu_r, 2\mu_f)$ and $(2\mu_r, 0.5\mu_f)$  will be discarded.

\item   \textbf{log correction}: Whether to include corrections due to the
  evolution of the strong coupling constant in the virtual
  corrections. Allowed values are \lstinline!true! and \lstinline!false!.

\item   \textbf{event output}: Specifies the name of a single event output
  file or a list of such files. The file format is either specified
  explicitly or derived from the suffix. For example, \lstinline!events.lhe!
  or, equivalently \lstinline!Les Houches: events.lhe! generates an output
  event file \lstinline!events.lhe! in the Les Houches format. The supported
  formats are

  \begin{itemize}
  \item \lstinline!file.lhe! or \lstinline!Les Houches: file!: The Les
    Houches event file format.
  \item \lstinline!file.hepmc! or \lstinline!HepMC: file!: The \text{HepMC} format.
  \end{itemize}

\item   \textbf{random generator}: Sets parameters for random number
  generation.
  \begin{itemize}
  \item \textbf{name}: Which random number generator to use. Currently,
    \lstinline!mixmax!~\cite{Savvidy:1988py,Savvidy:2014ana} and
    \lstinline!ranlux64!~\cite{Luscher:1993dy} are supported. See the CLHEP
    documentation~\cite{CLHEP} for details on the generators.

  \item \textbf{seed}: The seed for random generation. This should be a
    single number for \lstinline!mixmax! and the name of a state file for \lstinline!ranlux64!.
  \end{itemize}

\item \textbf{analysis}: Name and settings for the event analyses;
    either a custom analysis plug-in or \texttt{Rivet}. For the first the
    \lstinline!plugin!  sub-entry should be set to the analysis file
    path. All further entries are passed on to the analysis. To use
    \texttt{Rivet} a list of \texttt{Rivet} analyses have to be given in \lstinline!Rivet!
    and a prefix for the yoda file has to be set through
    \lstinline!output!. See section~\ref{sec:custom_analysis} for
    details.
\item   \textbf{Higgs coupling}: This collects a number of settings concerning
  the effective coupling of the Higgs boson to gluons. This is only
  relevant for the production process of a Higgs boson with jets and
  only supported if HEJ 2 was compiled with QCDLoop support.

  \begin{itemize}
  \item \textbf{use impact factors}: Whether to use impact factors for
    the coupling to the most forward and most backward partons. Impact
    factors imply the infinite top-quark mass limit.

  \item   \textbf{mt}: The value of the top-quark mass in GeV. If this is not
    specified, the limit of an infinite mass is taken.

  \item   \textbf{include bottom}: Whether to include the Higgs coupling to
    bottom quarks.

   \item  \textbf{mb}: The value of the bottom-quark mass in GeV.
  \end{itemize}
\end{itemize}


\section{The HEJ fixed order generator}
\label{sec:HEJFOG}

For high jet multiplicities event generation with standard fixed-order
generators becomes increasingly cumbersome. For example, the
leading-order production of a Higgs Boson with five or more jets is
computationally prohibitively expensive.

To this end, HEJ 2 provides the HEJFOG fixed-order generator
that allows to generate events with high jet multiplicities. To
facilitate the computation the limit of Multi-Regge Kinematics with
large invariant masses between all outgoing particles is assumed in the
matrix elements. The typical use of the HEJFOG is to supplement
low-multiplicity events from standard generators with high-multiplicity
events before using the HEJ 2 program to add high-energy
resummation.

\subsection{Installation}
\label{sec:HEJFOG_install}

The HEJFOG comes bundled together with HEJ 2 and the installation is
very similar. After downloading HEJ 2 and installing the prerequisites
as described in section~\ref{sec:prereq} the HEJFOG can be installed
with
\begin{lstlisting}
cmake /path/to/FixedOrderGen -DCMAKE_INSTALL_PREFIX=target/directory -DCMAKE_BUILD_TYPE=Release
make install
\end{lstlisting}
where \lstinline!/path/to/FixedOrderGen! refers to the \lstinline!FixedOrderGen!
subdirectory in the HEJ 2 directory.

If HEJ 2 was installed to a non-standard location, it may be necessary
to specify the directory containing \lstinline!HEJ-config.cmake!. If the
base installation directory is \lstinline!/path/to/HEJ!,
\lstinline!HEJ-config.cmake! should be found in
\lstinline!/path/to/HEJ/lib/cmake/HEJ! and the commands for installing
the HEJFOG would read

\begin{lstlisting}
cmake /path/to/FixedOrderGen -DHEJ_DIR=/path/to/HEJ/lib/cmake/HEJ -DCMAKE_INSTALL_PREFIX=target/directory
make install
\end{lstlisting}
The installation can be tested with::
\begin{lstlisting}
make test
\end{lstlisting}
provided that the CT10nlo PDF set is installed.

\subsection{Running the fixed-order generator}
\label{sec:HEJFOG_run}

After installing the HEJFOG you can modify the provided
configuration file \lstinline!configFO.yml! and run the generator with:
\begin{lstlisting}
  HEJFOG configFO.yml
\end{lstlisting}
The resulting event file, by default named \lstinline!HEJFO.lhe!, can then be
fed into HEJ 2 like any event file generated from a standard
fixed-order generator, see section~\ref{sec:start}.

\subsection{Settings}
\label{sec:HEJFOG_settings}

Similar to HEJ 2, the HEJFOG uses a YAML configuration file. The settings are

\begin{itemize}
\item \textbf{process}: The scattering process for which events are
  being generated. The format is \lstinline!in1 in2 => out1 out2 ...!.

  The incoming particles, \lstinline!in1!, \lstinline!in2! can be
  \begin{itemize}
  \item quarks: \lstinline!u!, \lstinline!d!, \lstinline!u_bar!, and
    so on,
  \item gluons: \lstinline!g!,
  \item protons \lstinline!p! or antiprotons \lstinline!p_bar!.
  \end{itemize}
  At most one of the outgoing particles can be a boson. At the moment
  only the Higgs boson \lstinline!h! is supported. All other outgoing
  particles are jets. Multiple jets can be grouped together, so
  \lstinline!p p => h j j! is the same as \lstinline!p p => h 2j!.
  There have to be at least two jets.

\item \textbf{events}: Specifies the number of events to generate.

\item \textbf{jets}: Defines the properties of the generated jets.

  \begin{itemize}
  \item \textbf{min pt}: Minimum jet transverse momentum in GeV.

  \item \textbf{peak pt}: Optional setting to specify the dominant jet
    transverse momentum in GeV. If the generated events are used as
    input for HEJ resummation, this should be set to the minimum
    transverse momentum of the resummation jets. The effect is that only
    a small fraction of jets will be generated with a transverse
    momentum below the value of this setting.

  \item \textbf{algorithm}: The algorithm used to define jets. Allowed
    settings are \lstinline!kt!, \lstinline!cambridge!, \lstinline!antikt!, \lstinline!cambridge for passive!.
    See the FastJet documentation for a description of these algorithms.

  \item \textbf{R}: The R parameter used in the jet algorithm.
  \item \textbf{max rapidity}: Maximum absolute value of the jet
    rapidity.
  \end{itemize}

\item \textbf{beam}: Defines various properties of the collider beam.

  \begin{itemize}
  \item \textbf{energy}: The beam energy in GeV. For example, the 13 TeV LHC
    corresponds to a value of 6500.

  \item \textbf{particles}: A list \lstinline![p1, p2]! of two beam
    particles. The only supported entries are protons \lstinline!p! and
    antiprotons \lstinline!p_bar!.
  \end{itemize}

  \item \textbf{pdf}: The LHAPDF number of the PDF set. For example, 230000
  corresponds to an NNPDF 2.3 NLO PDF set.

\item \textbf{subleading fraction}:
   This setting is related to the fraction of events that are not FKL
   configurations and thus subleading in the high-energy
   limit. Currently only unordered emissions are implemented, and only
   for Higgs boson plus multijet processes. This value must be positive
   and not larger than 1. It should typically be chosen between 0.01 and
   0.1. Note that while this parameter influences the chance of
   generating subleading configurations, it generally does not
   correspond to the actual fraction of subleading events.

\item \textbf{subleading channels}:
   Optional parameter to select the production of specific channels that
   are subleading in the high-energy limit. Only has an effect if
   \lstinline!subleading fraction! is non-zero. Currently three values are
   supported:

   \begin{itemize}
   \item \lstinline!all!: All subleading channels are allowed. This is
     the default.
   \item \lstinline!none!: No subleading contribution, only
     FKL configurations are allowed. This is equivalent to
     \lstinline!subleading fraction: 0!.
   \item \lstinline!unordered!: Unordered emission are allowed.

     Unordered emission are any rapidity ordering where exactly one
     gluon is emitted outside the FKL rapidity ordering. More precisely,
     if at least one of the incoming particles is a quark or antiquark
     and there are more than two jets in the final state,
     \lstinline!subleading fraction! states the probability that the
     flavours of the outgoing particles are assigned in such a way that
     an unordered configuration arises.
   \end{itemize}

 \item \textbf{unweight}: This setting defines the parameters for the partial
   unweighting of events. You can disable unweighting by removing this
   entry from the configuration file.

  \begin{itemize}
  \item \textbf{sample size}: The number of weighted events used
    to calibrate the unweighting.  A good default is to set this to the
    number of target events. If the number of events is large this
    can lead to significant memory consumption and a lower value should
    be chosen. Contrarily, for large multiplicities the unweighting
    efficiency becomes worse and the sample size should be increased.

  \item \textbf{max deviation}: Controls the range of events to which
    unweighting is applied. A larger value means that a larger fraction
    of events are unweighted.  Typical values are between -1 and 1.
  \end{itemize}

\item \textbf{Particle properties}: Specifies various properties of the
  different particles (Higgs, W or Z).  This is only relevant if the
  chosen process is the production of the corresponding particles with
  jets. For example, for the process \lstinline!p p => h 2j! the
  \lstinline!mass!, \lstinline!width! and (optionally) \lstinline!decays!
  of the \lstinline!Higgs! boson are required, while all other particle
  properties will be ignored. In the current version, the production of
  W and Z bosons is not implemented and those entries will always be
  ignored. This will change in future versions.
  \begin{itemize}
  \item \textbf{Higgs}, \textbf{W+}, \textbf{W-} or \textbf{Z}:
     The particle (Higgs, W$^+$, W$^-$, Z) for which the following
     properties are defined.

    \begin{itemize}
      \item \textbf{mass}: The mass of the particle in GeV.

      \item \textbf{width}: The total decay width of the particle in GeV.

      \item \textbf{decays}:
      Optional setting specifying the decays of the particle. Only the decay
      into two particles is implemented. Each decay has the form
      \lstinline!{into: [p1,p2], branching ratio: r}!
      where \lstinline!p1! and \lstinline!p2! are the particle names of the
      decay product (e.g. \lstinline!photon!) and \lstinline!r! is the branching
      ratio.
      Decays of a Higgs boson are treated as the production and subsequent
      decay of an on-shell Higgs boson, so decays into e.g. Z bosons are not
      supported.
    \end{itemize}
  \end{itemize}
\item \textbf{scales}: Specifies the renormalisation and factorisation
  scales for the output events. For details, see the corresponding entry
  in section~\ref{sec:settings}. Note that this should usually be a
  single value, as the weights resulting from additional scale choices
  will simply be ignored when adding high-energy resummation with HEJ 2.

\item \textbf{event output}: Specifies the name of a single event output
  file or a list of such files. See the corresponding entry in section~\ref{sec:settings} for details.

\item \textbf{random generator}: Sets parameters for random number
  generation. See section~\ref{sec:settings} for details.

\item \textbf{analysis}: Specifies the name and settings for a custom
  analysis library. This can be useful to specify cuts at the
  fixed-order level. See the corresponding entry in section~\ref{sec:settings} for details.

\item \textbf{Higgs coupling}: This collects a number of settings concerning
  the effective coupling of the Higgs boson to gluons. See the
  corresponding entry in section~\ref{sec:settings} for details

\end{itemize}

\section{Writing custom analyses}
\label{sec:custom_analysis}

HEJ 2 and the HEJ fixed-order generator can generate \texttt{HepMC} files, so it
is always possible to run a \texttt{Rivet} analysis on these. However if HEJ 2
was compiled with \texttt{Rivet} support one can use the native \texttt{Rivet}
interface. For example
\begin{lstlisting}
analysis:
  rivet: [MC_XS, MC_JETS]
  output: HEJ
\end{lstlisting}
would call the generic \lstinline!MC_XS! and \lstinline!MC_JETS! \texttt{Rivet}
analyses and write the result into \lstinline!HEJ[.Scalename].yoda!. HEJ 2
will then run \texttt{Rivet} over all different scales separately and write out
each into a different yoda file. Alternatively instead of using \texttt{Rivet},
one can provide a custom analysis inside a C++ library.

An analysis is a class that derives from the abstract \lstinline!Analysis!
base class provided by HEJ 2. It has to implement three public
functions:
\begin{itemize}
\item  The \lstinline!pass_cuts! member function return true if and
  only if the given event (first argument) passes the analysis cuts.

\item  The \lstinline!fill! member function adds an event to the analysis,
  which for example can be used to fill histograms. HEJ 2 will only pass
  events for which \lstinline!pass_cuts! has returned true.

\item  The \lstinline!finalise! member function is called after all events
  have been processed. It can be used, for example, to print out or save
  the analysis results.
\end{itemize}
The \lstinline!pass_cuts! and \lstinline!fill! functions take two arguments: the
resummation event generated by HEJ 2 and the original fixed-order
input event. Usually, the second argument can be ignored. It can be
used, for example, for implementing cuts that depend on the ratio of the
weights between the fixed-order and the resummation event.

In addition to the two member functions, there has to be a global
\lstinline!make_analysis! function that takes the analysis parameters in the form of
a YAML \lstinline!Node! and returns a \lstinline!std::unique_ptr! to the
Analysis.

The following code creates the simplest conceivable analysis.
\begin{lstlisting}[language=C++]
#include <memory> // for std::unique_ptr

#include "HEJ/Analysis.hh"

class MyAnalysis: public HEJ::Analysis {
  public:
  MyAnalysis(YAML::Node const & /* config */) {}

  void fill(
    HEJ::Event const & /* event */,
    HEJ::Event const & /* FO_event */
  ) override { }

  bool pass_cuts(
    HEJ::Event const & /* event */,
    HEJ::Event const & /* FO_event */
  ) override {
      return true;
  }

  void finalise() override { }

};

extern "C"
std::unique_ptr<HEJ::Analysis> make_analysis(
  YAML::Node const & config
){
    return std::make_unique<MyAnalysis>(config);
}
\end{lstlisting}
After saving this code to a file, for example \lstinline!myanalysis.cc!,
this code can be compiled into a shared library. Using the \lstinline!g++! compiler, the
library can be built with
\begin{lstlisting}
g++ $(HEJ-config --cxxflags) -fPIC -shared -Wl,-soname,libmyanalysis.so -o libmyanalysis.so myanalysis.cc
\end{lstlisting}
With \lstinline!g++! it is also good practice to add
\lstinline!__attribute__((visibility("default")))! after \lstinline!extern "C"!
in the above code snippet and then compile with the additional flag
\lstinline!-fvisibility=hidden! to prevent name clashes.

The analysis can be used in HEJ 2 or the HEJ fixed-order generator by
adding
\begin{lstlisting}
analysis:
  plugin: /path/to/libmyanalysis.so
\end{lstlisting}
to the .yml configuration file.

As a more interesting example, here is the code for an analysis that
sums up the total cross section and prints the result to both standard
output and a file specified in the .yml config with

\begin{lstlisting}
analysis:
   plugin: analysis/build/directory/src/libmy_analysis.so
   output: outfile
\end{lstlisting}
To access the configuration at run time, HEJ 2 uses the yaml-cpp
library. The analysis code itself is
\begin{lstlisting}[language=C++]
#include <memory>
#include <iostream>
#include <fstream>
#include <string>
#include <cmath>

#include "HEJ/Analysis.hh"
#include "HEJ/Event.hh"

#include "yaml-cpp/yaml.h"

class MyAnalysis: public HEJ::Analysis {
public:
  MyAnalysis(YAML::Node const & config):
  xsection_{0.}, xsection_error_{0.},
  outfile_{config["output"].as<std::string>()}
  {}

  void fill(
    HEJ::Event const & event,
    HEJ::Event const & /* FO_event */
  ) override {
    const double wt = event.central().weight;
    xsection_ += wt;
    xsection_error_ += wt*wt;
  }

  bool pass_cuts(
    HEJ::Event const & /* event */,
    HEJ::Event const & /* FO_event */
  ) override {
    return true;
  }

  void finalise() override {
    std::cout << "cross section: " << xsection_ << " +- "
      << std::sqrt(xsection_error_) << "\n";
    std::ofstream fout{outfile_};
    fout << "cross section: " << xsection_ << " +- "
      << std::sqrt(xsection_error_) << "\n";
  }

  private:
  double xsection_, xsection_error_;
  std::string outfile_;

};

extern "C"
std::unique_ptr<HEJ::Analysis> make_analysis(
    YAML::Node const & config
){
  return std::make_unique<MyAnalysis>(config);
}
\end{lstlisting}


\section{Custom scales}
\label{sec:scales}

HEJ 2 comes with a small selection of built-in renormalisation and
factorisation scales, as described in section~\ref{sec:settings}. In
addition to this, user-defined scales can be imported from custom
libraries.

\subsection{Writing the library}
\label{sec:scale_library}

Custom scales are defined through C++ functions that take an event and
compute the corresponding scale. As an example, let us consider a
function returning the transverse momentum of the softest jet in an
event. To make it accessible from HEJ 2, we have to prevent C++
name mangling with \lstinline!extern "C"!:
\begin{lstlisting}[language=C++]
#include "HEJ/Event.hh"

extern "C"
double softest_jet_pt(HEJ::Event const & ev){
  const auto softest_jet = sorted_by_pt(ev.jets()).back();
  return softest_jet.perp();
}
\end{lstlisting}
After saving this code to some file \lstinline!myscales.cc!, we can compile
it to a shared library. With the \lstinline!g++! compiler this can be done
with the command
\begin{lstlisting}
g++ $(HEJ-config --cxxflags) -fPIC -shared -Wl,-soname,libmyscales.so -o libmyscales.so myscales.cc
\end{lstlisting}

\subsection{Importing the scale into HEJ 2}
\label{sec:scale_import}

Our custom scale can now be imported into HEJ 2 by adding the following
lines to the YAML configuration file
\begin{lstlisting}
import scales:
  /path/to/libmyscales.so: softest_jet_pt
\end{lstlisting}
It is also possible to import several scales from one or more libraries:
\begin{lstlisting}
import scales:
   /path/to/libmyscales1.so: [first_scale, second_scale]
   /path/to/libmyscales2.so: [another_scale, yet_another_scale]
\end{lstlisting}
The custom scales can then be used as usual in the \textbf{scales} setting, for example
\begin{lstlisting}
scales: [H_T, softest_jet_pt, 2*softest_jet_pt]
\end{lstlisting}


\section{Using HEJ 2 as a library}
\label{sec:lib_use}

As mentioned before, HEJ 2 can also be used as a library, which allows
lots of flexibility. The documentation of the complete functionality can
be found on \url{https://hej.web.cern.ch/HEJ/doc/2.0/library}.

As an example, we show a toy program that computes the square of a
matrix element in the \HEJ{} approximation for a single event. First, we
include the necessary header files:
\begin{lstlisting}[language=C++]
#include "HEJ/Event.hh"
#include "HEJ/MatrixElement.hh"
\end{lstlisting}
We then specify the incoming and outgoing particles. A particle
has a type and four-momentum $(p_x, p_y, p_z, E)$. For instance, an
incoming gluon could be defined as
\begin{lstlisting}[language=C++]
fastjet::PseudoJet momentum{0, 0, 308., 308.};
HEJ::Particle gluon_in{HEJ::ParticleID::gluon, momentum};
\end{lstlisting}
We collect all incoming and outgoing particles in a partonic event. Here
is an example for a partonic $gu \to gghu$ event:
\begin{lstlisting}[language=C++]
HEJ::UnclusteredEvent partonic_event;

// incoming particles
partonic_event.incoming[0] = {
  HEJ::ParticleID::gluon,
  { 0., 0., 308., 308.}
};
partonic_event.incoming[1] = {
  HEJ::ParticleID::up,
  { 0., 0.,-164., 164.}
};
// outgoing particles
partonic_event.outgoing.push_back({
  HEJ::ParticleID::higgs,
  { 98., 82., 14., 180.}
});
partonic_event.outgoing.push_back({
  HEJ::ParticleID::up,
  { 68.,-54., 36.,  94.}
});
partonic_event.outgoing.push_back({
  HEJ::ParticleID::gluon,
  {-72.,  9., 48.,  87.}
});
partonic_event.outgoing.push_back({
  HEJ::ParticleID::gluon,
  {-94.,-37., 46., 111.}
});
\end{lstlisting}
Alternatively, we could read the event from a Les Houches event file,
possibly compressed with \texttt{gzip}. For this, the additional header
files \lstinline!HEJ/stream.hh! and \lstinline!LHEF/LHEF.h! have to be
included.
\begin{lstlisting}[language=C++]
HEJ::istream in{"events.lhe.gz"};
LHEF::Reader reader{in};
reader.readEvent();
HEJ::UnclusteredEvent partonic_event{reader.hepeup};
\end{lstlisting}

In this specific example we will later choose a constant value for the
strong coupling, so that the HEJ matrix element does not depend on the
renormalisation scale. However, in a more general scenario, we will want
to set a central scale:
\begin{lstlisting}[language=C++]
partonic_event.central.mur = 50.;
\end{lstlisting}
It is possible to add more scales in order to perform scale variation:
\begin{lstlisting}[language=C++]
partonic_event.variations.resize(2);
partonic_event.variations[0].mur = 25.;
partonic_event.variations[1].mur = 100.;
\end{lstlisting}

In the next step, we leverage FastJet to construct an event with
clustered jets. Here, we use anti-$k_t$ jets with $R=0.4$ and transverse
momenta of at least $30\,$GeV.
\begin{lstlisting}[language=C++]
const fastjet::JetDefinition jet_def{
 fastjet::JetAlgorithm::antikt_algorithm, 0.4
};
const double min_jet_pt = 30.;
HEJ::Event event{partonic_event, jet_def, min_jet_pt};
\end{lstlisting}
In order to calculate the Matrix element, we now have to fix the physics
parameters. For the sake of simplicity, we assume an effective coupling
of the Higgs boson to gluons in the limit of an infinite top-quark mass
and a fixed value of $\alpha_s = 0.118$ for the strong coupling.
\begin{lstlisting}[language=C++]
const auto alpha_s = [](double /* mu_r */) { return 0.118; };
HEJ::MatrixElementConfig ME_config;
// whether to include corrections from the
// evolution of \alpha_s in virtual corrections
ME_config.log_correction = false;
HEJ::MatrixElement ME{alpha_s, ME_config};
\end{lstlisting}
If QCDLoop is installed, we can also take into account the full loop
effects with finite top and bottom quark masses:
\begin{lstlisting}[language=C++]
HEJ::MatrixElementConfig ME_config;
ME_config.Higgs_coupling.use_impact_factors = false;
ME_config.Higgs_coupling.mt = 163;
ME_config.Higgs_coupling.include_bottom = true;
ME_config.Higgs_coupling.mb = 2.8;
\end{lstlisting}
Finally, we can compute and print the square of the matrix element with
\begin{lstlisting}[language=C++]
std::cout << "HEJ ME: " << ME(event).central << '\n';
\end{lstlisting}
In the case of scale variation, the weight associated with the scale
\lstinline!event.variations[i].mur! is
\lstinline!ME(event).variations[i]!.

Collecting the above pieces, we have the following program:
\begin{lstlisting}[language=C++]
#include "HEJ/Event.hh"
#include "HEJ/MatrixElement.hh"

int main(){
  HEJ::UnclusteredEvent partonic_event;
  // incoming particles
  partonic_event.incoming[0] = {
    HEJ::ParticleID::gluon,
    { 0., 0., 308., 308.}
  };
  partonic_event.incoming[1] = {
    HEJ::ParticleID::up,
    { 0., 0.,-164., 164.}
  };
  // outgoing particles
  partonic_event.outgoing.push_back({
    HEJ::ParticleID::higgs,
    { 98., 82., 14., 180.}
  });
  partonic_event.outgoing.push_back({
    HEJ::ParticleID::up,
    { 68.,-54., 36.,  94.}
  });
  partonic_event.outgoing.push_back({
    HEJ::ParticleID::gluon,
    {-72.,  9., 48.,  87.}
  });
  partonic_event.outgoing.push_back({
    HEJ::ParticleID::gluon,
    {-94.,-37., 46., 111.}
  });

  const fastjet::JetDefinition jet_def{
    fastjet::JetAlgorithm::antikt_algorithm, 0.4
  };
  const double min_jet_pt = 30.;
  HEJ::Event event{partonic_event, jet_def, min_jet_pt};

  const auto alpha_s = [](double /* mu_r */) { return 0.118; };
  HEJ::MatrixElementConfig ME_config;
  // whether to include corrections from the
  // evolution of \alpha_s in virtual corrections
  ME_config.log_correction = false;
  HEJ::MatrixElement ME{alpha_s, ME_config};

  std::cout
    << "HEJ ME: " << ME(event).central
    << " = tree * virtual = " << ME.tree(event).central
    << " * " << ME.virtual_corrections(event).central
    << '\n';
}
\end{lstlisting}
After saving the above code to a file \lstinline!matrix_element.cc!, it
can be compiled into an executable \lstinline!matrix_element! with a
suitable compiler. For example, with \lstinline!g++! this can be done
with the command
\begin{lstlisting}[language=sh]
g++ -o matrix_element matrix_element.cc -lHEJ -lfastjet
\end{lstlisting}
If HEJ or any of the required libraries (see section~\ref{sec:prereq})
was installed to a non-standard location, it may be necessary to
explicitly specify the paths to the required header and library
files. This can be done with the \lstinline!HEJ-config! executable and
similar programs for the other dependencies:
\begin{lstlisting}[language=sh]
g++ $(fastjet-config --cxxflags) $(HEJ-config --cxxflags) -o matrix_element matrix_element.cc $(HEJ-config --libs) $(fastjet-config --libs)
\end{lstlisting}


\section{Summary}
\label{sec:summary}

We have presented the HEJ~2 event generator which may be used to generate Monte
Carlo events for hadron colliders at leading-logarithmic (LL) accuracy in
$\hat{s}/p_\perp^2$, for multi-jet production and for Higgs boson production in
association with jets.  It takes as input fixed-order samples, currrently at
leading-order (LO), and maintains this accuracy to give combined LO+LL
predictions.  The addition of the LL terms has been seen to be particularly
significant in regions of large invariant mass between jets (or equivalently
large rapidity separation), which is particularly pertinent for gluon-fusion
production of a Higgs boson in association with dijets.

The HEJ~2 code is publicly available from \url{https://hej.web.cern.ch}.  In
this contribution we have outlined all necessary details to download, install
and run HEJ~2, including a full description of the possible settings which can
be user-defined in an input file.  We have further discussed how to create a
general analysis of the events produced and how HEJ~2 can also be used as a
standalone library.  Therefore, this documentation will allow anyone to generate
their own predictions for arbitrary experimental setups at the LHC and future
colliders.


\section*{Acknowledgements}

The authors would like to thank Gavin Salam for discussions on the
anti-$k_t$ jet clustering algorithm~\cite{Cacciari:2008gp}.

This work has received funding from the European Union’s Horizon 2020
research and innovation programme as part of the Marie Skłodowska-Curie
Innovative Training Network MCnetITN3 (grant agreement no. 722104), the
Marie Skłodowska-Curie grant agreement No. 764850, SAGEX, and COST
action CA16201: “Unraveling new physics at the LHC through the precision
frontier”, and from the UK Science and Technology Facilities Council
(STFC). JMS is supported by a Royal Society University Research
Fellowship and the ERC Starting Grant 715049 “QCDforfuture”.






\bibliographystyle{elsarticle-num}
\bibliography{papers}

\end{document}